\documentclass[12pt,preprint]{aastex}
\usepackage{latexsym,graphicx,rotating}

\newcommand{\beq}{\begin{equation}}
\newcommand{\eeq}{\end{equation}}

\begin{document}
\bibliographystyle{icarus}

\title{Near-Infrared Photometry of Irregular Satellites of Jupiter and Saturn}

\shorttitle{NIR Photometry of Irregular Satellites}
\shortauthors{Grav, Holman}
\medskip

\author{Tommy~Grav\altaffilmark{1}}
\affil{\footnotesize \it Institute of Theoretical Astrophysics, University in Oslo,\\
	Postbox 1029 Blindern, 0359 Oslo, Norway (tommy.grav@astro.uio.no) \\
	 \& \\
	Harvard-Smithsonian Center for Astrophysics, \\ 
	MS51, 60 Garden Street, Cambridge MA 02138}
\email{tgrav@cfa.harvard.edu}
\author{Matthew~J.~Holman}
\affil{\footnotesize \it Harvard-Smithsonian Center for Astrophysics, \\
MS51, 60 Garden Street, Cambridge, MA 02138}
\email{mholman@cfa.harvard.edu}
\vspace{1cm}
\altaffiltext{1}{Visiting Astronomer, Gemini North Oberservatory}


\begin{abstract}
We present $JHK_s$ photometry of 10 Jovian and 4 Saturnian irregular satellites, taken with the Near-InfraRed Imager (NIRI) at the 8-m Gemini North Observatory on Mauna Kea, Hawaii. The observed objects have near-infrared colors consistent with C, P and D-type asteroids, although J~XII~Ananke and S~IX~Phoebe show weak indications of possible water features in the $H$ filter. The four members of the Himalia-family have similar near-infrared colors, as do the two members of the Gallic family, S~XX~Paaliaq and S~XXIX~Siarnaq.  From low resolution normalized reflectance spectra based on the broadband colors and covering $0.4\mu$m to $2.2\mu$m, the irregular satellites are identified as C-type (J~VII~Pasiphae, J~VI~Himalia and S~IX~Phoebe), P-type (J~XII~Ananke and J~XVIII~Themisto) and D-type (J~IX~Carme and J~X~Sinope), showing a diversity of origins of these objects.

\end{abstract}
\section{Introduction}

The satellites of the giant planets can be divided into two classes, regular and irregular, based on their orbital characteristics. The regular satellites are thought to have been formed in a planetocentric disk with nearly circular, prograde orbits close to the equatorial plane of the planet. The irregular satellites, however, have highly inclined and eccentric orbits, both prograde and retrograde. These orbital characteristics suggest that they formed outside the circumplanetary disk and were subsequently captured. Objects in heliocentric orbits can be temporarily captured,  for $10-100$ orbits. Several mechanisms for making a temporary capture permanent have been proposed: 1)  fragmentation during a single collision between an outer satellite and an asteroid \citep{Colombo.1971}; 2) a rapid increase in the mass of a planet during formation;  and 3)  gas drag from an extended gas envelope \citep{Pollack.1979} or a flattened disk \citep{Cuk.2003} around the planet. \cite{Astakhov.2003} suggest that chaos-assisted capture through dynamical diffusion with a flattened disk could dominate in regions close to the Kozai and other secular resonances. All these theories imply different dynamical distributions, capture time scales, and satellite ages. 

Determining the physical characteristics of the irregular satellites is key to understanding their origin. In \cite{Grav.2003b} and \cite{Grav.2003dps} we reported a large set of optical $BVRI$ colors of the irregular satellites and showed that most of the dymanical clusters have homogeneous colors, indicating that the clusters are the result of the capture of larger progenitors that were subsequently disrupted \citep{Gladman.2001}.  The optical colors are consistent with the progenitors being outer main belt asteroids due to the lack of the extremely {\it red} colors found among the Centaur and Trans-Neptunian object (TNO) population. Extending the wavelength coverage into the near-infrared significantly helps in distinguishing between the possible origins of the progenitors. 

\citet{Cruikshank.1980} and \citet{Degewij.1980a} reported near-infrared observations of the irregular satellites J~VI~Himalia and S~IX~Phoebe and found that the colors were different from those of the regular satellites, looking like low albedo carbonaceous chondritic asteroids. $IJHK$ band spectroscopy of J~VI~Himalia was collected by \citet{Dumas.1998} and \citet{Brown.2000} presented $1.3-2.4\mu$m near-infrared spectra of J~VI~Himalia, J~VII~Elara, J~VII~Pasiphae, S~IX~Phoebe and N~II~Nereid. The three Jovian irregular satellites have flat, featureless spectra consistent with dark, carbonaceous chondrities. Both S~IX~Phoebe and N~II~Nereid had broad absorption features centered on $2\mu$m, indicating the presence of water ice. They argued that the presence of water ice on S~IX~Phoebe demonstrated that it is a captured outer solar system body, rather than an asteroid. The Two-Micron All Sky Survey (2MASS) detected 6 of the then 8 known Jovian irregular satellites in the  $JHK_s$ bandpasses \citep{Sykes.2000}. They found colors that were consistent with C- and D/P-type asteroids, except for J~X~Sinope, which had an extreme $J-H$ color. Their $1\sigma$ errors were significant, however, evenly distributed in the range from $0.03$ (J~VI~Himalia) to $0.28$ (J~X~Sinope). 

This work reports near-infrared photometry of 10 Jovian and 4 Saturnian irregular satellites. In section 2 we discuss the observations and the routines used for image reduction, data analysis and calibration. In section 3 we discuss our result and compare it to the results of other observations of irregular satellites in the near-infrared wavelength region. In section 4 we discuss the implications of our observations.
Note that all use of C-, P- and D-type asteroids in the following text referees to the taxonomical classification scheme developed by \citet{Tholen.1984}.

\section{The Observations}
Using the Gemini North 8m Telescope with the NIRI (Near InfraRed Imager) \citep{Hora.1995} we observed the irregular satellites in classical observing mode on Feb. 17-19th, 2003. The camera was used in its $f/6$ direct imaging configuration, giving it a scale of $0.117$ arcsec per pixels. All three nights were clear and photometric, with seeing of $0.6$ to $1.3$ arcseconds. We used a standard $JHK_s$ filter set, where $K_s$ (or $K$ ``short'') is a medium band modified $K$ filter used to reduce thermal background from the warm telescope and surrounding structures \citep{Persson.1998}. The $K_s$ and $K$ filter have virtually identical throughput with differences on the $\pm 0.01$ magnitude level \citep{Persson.1998}. 

The data reduction was performed in IRAF \citep{Tody.1986,Tody.1993} using the Gemini and DAOPHOT \citep{Stetson.1990} packages. Flatfield and dark exposures were taken at morning and/or evening twilight and were combined to make normalized flatfield images and bad-pixel maps. Skyframes were created from the dithered science image. The science images were skyframe subtracted and flatfield divided, resulting in images with less than $2$\% variation across the field. Some unidentified $60$Hz  noise coming from the instrument or telescope is evident in some of the images but is too low to significantly contribute to the estimated errors of the photometry.

Aperture photometry was performed. The signal-to-noise ratio for the observed targets was large enough that aperture correction photometry was not needed. An aperture of $\sim 3.0$ arcseconds (25 pixels) was used, while a sky annulus starting at $\sim 3.6$ arcseconds (30 pixels) with a width of $\sim 2.4$ arcseconds (20 pixels wide).

A number of UKIRT near infrared standard stars \citep{UKIRT.standards} were taken throughout the night spanning the airmass of our science targets. The standard stars were used to derive zero points, airmass corrections and filter color corrections.

\section{Results}

The results of our observations of ten Jovian and four Saturnian irregular satellites are given in Table \ref{tab:obs}. Seven of these satellites were previously unobserved in the $JHK_s$ bandpasses. Our results of the six brightest Jovian irregular satellites are generally consistent with \citet{Sykes.2000}, but we have significantly improved on the errors of the photometry (the average of  the  $JHK_s$ color errors reported here is $\sim 0.03$). Figure \ref{fig:jhk} shows the J-H vs. H-K colors of the objects observed. On this figure we have also plotted the areas of $JHK_s$ phase space populated by known C, D, F and P-type asteroids as observed by the Two Micron All Sky Survey \citep[2MASS;][]{Sykes.2000b}.  From Figure \ref{fig:jhk} we see that most of the satellites are consistent with these outer main belt asteroids. There are however three satellites, J~VII~Elara, J~XII~Ananke and S~IX~Phoebe, that lie outside these areas. It should be noted that the few Centaurs observed in the near-infrared also fall into a similar range of near-infrared colors \citep{Weintraub.1997,Brown.2000}.

Figures \ref{fig:jup_spectra}-\ref{fig:sat_spectra} shows the low resolution spectra of the objects observed. The spectra shown are normalized at the $V$ filter ($0.53 \mu$m). The BVRI colors have been taken from \citet{Rettig.2001} and \citet{Grav.2003b} and solar colors were adopted as $B-V=0.67$,$V-R=0.36$, $V-I=0.71$, $V-J=1.08$, $V-H=1.37$ and $V-K=1.43$ \citep{Jewitt.1998,Degewij.1980a}. Note that there are values given for solar H-K and J-K colors by other authors that differ from this \citep[for example ][]{Johnson.1975}.

Only a few V-J colors of irregular satellites are available. Values for J~VI~Himalia and S~IX~Phoebe were taken from \cite{Cruikshank.1980} and \cite{Degewij.1980a}. To determine values for $V-J$ for the other targets the $V$ observations from \citet{Grav.2003b} were corrected for the heliocentric, geocentric and phase angle difference between the $V$ and $J$ observations using
\beq
	H(1,1,0) = m - 5log(\Delta R) -  \beta \alpha
\eeq
where $m$ is the apparent magnitude, $R$ is the heliocentric distance, $\Delta$ is the geocentric distance, $\beta$ is a constant phase coefficient and $\alpha$ is the phase angle. S~IX~Phoebe is, however the only irregular satellite of Jupiter and Saturn for which a phase coefficient has been determined \citep{Kruse.1986}. At low phase angles ($\alpha < 1^\circ$) they found  $\beta = 0.18 \pm 0.04$ mag/${}^\circ$, but at larger phase angles ($\alpha > 2$) it falls to $\beta \sim 0.10$ mag/${}^\circ$ \citet{Kruse.1986}. We have adopted the value of $\beta = 0.10$ mag/${}^\circ$ since all of the observations from \citet{Grav.2003b} are at higher phase angles than $2^\circ$. Note that no correction due to rotational variations has been applied. To better compare members of the same families the $V-J$ values of the brightest member were used for all the family members. This of course makes the spectra seem more similar at a first glance than might be justified, but it significantly increases our ability to compare the visual and infrared observations. The errors plotted for the $JHK_s$ colors do not take into account the errors in our derivation of the $V-J$ colors, since this does not change any of our arguments. 
 
For the Himalia and Gallic (the Saturnian $45^\circ$-inclination) families we observed more than one member and thus can test if the homogeneous colors found in \citet{Grav.2003b} extend into the near- infrared. Figure \ref{fig:jup_spectra} shows the low resolution spectra for the four members of the Himalia-family (S/2000 J11 has not been observed beyond the discovery apparition and is considered lost). The four spectra are very similar, all consistent with spectra of C-type asteroids. The only noticeable difference lies in the slightly redder color of J~XI~Lysithea in the visual wavelengths, but these differences are within the $3 \sigma$ errors. The low resolution spectra of the two objects observed from the Gallic family, S~XX~Paaliaq and S~XXIX~Siarnaq, are plotted in Figure \ref{fig:sat_spectra}. The two spectra are very similar and seem to be consistent with a P-type asteroidal surface. The results from this survey further strengthen the arguments from \citep{Grav.2003b} for the dynamical families being remainders of larger progenitors captured and subsequently fragmented due to collisions. 

Using the low resolution reflectance spectra we identify J~VIII~Pasiphae as a C-type asteroid, J~XII~Ananke and J~XVIII~Themisto as P-type asteroids, while J~IX~Carme, and  J~X~Sinope and J~XVII Callirrhoe are identified as D-type asteroids \cite[for examples of the different spectral classes see][]{Tholen.1989}. For the Saturnian irregular satellites S~IX~Phoebe is identified as having a F-type asteroidal surface, while the surface of S~XXVI~Albiorix is consistent with that of a P-type. The low resolution spectra show the large diversity among the irregular satellites of Jupiter and Saturn, ranging from the slightly blue spectrum of the Himalia-family members and S~IX~Phoebe, to the red spectrum of J~IX~Carme, J~X~Sinope and S~XXVI~Albiorix. Higher resolution spectra are needed however to secure these identifications. We looked for correlations between spectral type and orbital parameters, but no apparent patterns were found.

The data seem to indicate that S~IX~Phoebe and J~XII~Ananke have some small degree of contamination of water ices. A broad absorption feature centered on $\sim 2.0 \mu$m due to water ices has been detected on S~IX~Phoebe in spectra by \citet{Owen.1999} and \citet{Brown.2000}. High resolution near-infrared spectra are needed to confirm this feature on J~XII~Ananke. The extreme values of the $J-H$ and $H-K_s$ colors of J~IX~Sinope reported by \citet{Sykes.2000} were not seen in our data. The data are however consistent with each other on a $3\sigma$ level. It is conceivable that J~IX~Sinope does not have a homogeneous surface composition and that the two observations were performed at different rotational phases. 

Our observations of S~IX~Phoebe confirm the colors found by both \citet{Cruikshank.1980} and \citet{Degewij.1980a}. Most likely the two observations by \citet{Degewij.1980a}, that do not agree with our results, are taken at different rotational phases than our observations. S~IX~Phoebe is known to have a bright spot with a peak albedo of $0.11$  which is significantly higher than the $0.07$ albedo of the rest of the surface \citep{Simonelli.1999}.

\section{Conclusions}

We have collected near-infrared colors of 10 Jovian and 4 Saturnian irregular satellites. The colors are found to be consistent with the C,F,D and P type asteroids found in the outer parts of the main asteroid belt. The near-infrared colors show that four members of the Himalia-family have homogeneous colors, with weighted mean colors of $J-H=0.34\pm0.02$, $H-K_s=0.27\pm0.02$ and $J-K_s=0.65\pm0.02$. The the observed members of the Saturnian Gallic-family are also found to have homogeneous colors, with weighted mean colors of $J-H=0.37\pm0.01$, $H-K_s=0.16\pm0.02$ and $J-K_s=0.53\pm0.01$. The homogeneous colors of these two families further supports the hypothesis that these families are the remnants of larger progenitors that were captured and subsequently fragmented
\citep{Gladman.2001,Grav.2003b}.

Adding observations in the visual wavelengths and plotting the results as a reflectance spectra normalized at $V$ shows that the irregular satellites of Jupiter and Saturn have surfaces ranging from the neutral members of the Himalia-family and S~IX~Phoebe, to the red spectra of J~IX~Carme, J~X~Sinope and S~XXVI~Albiorix. Furthermore the low resolution spectra of J~XII~Ananke and S~IX~Phoebe show weak indications of broad absorption features centered on $2.0\mu$m from water ices. These features have been identified in S~IX~Phoebe using higher resolution spectra \cite{Owen.1999,Brown.2000}.

\section{Acknowledgments}

We would like to thank J. Jensen, S. Chan and the staff of the Gemini North Telescope for their excellent help in the taking and reduction of these onbservations. Based on observations obtained at the Gemini Observatory, which is operated by the Association of Universities for Research in Astronomy, Inc., under a cooperative agreement with the NSF on behalf of the Gemini partnership: the National Science Foundation (United States), the Particle Physics and Astronomy Research Council (United Kingdom), the National Research Council (Canada), CONICYT (Chile), the Australian Research Council (Australia), CNPq (Brazil), and CONICET (Argentina).

Tommy Grav is a Smithsonian Astrophysical Observatory Pre-doctoral Fellow at the Harvard-Smithsonian Center for Astrophysics, Cambridge, USA. 

This work was supported by NASA grants NAG5-9678 and NAG5-10438.

\newpage

\section{Table and Figure Captions}

{\bf Table 1.} This table shows the circumstances and  results of our observations.  Heliocentric ($r$) and geocentric ($\Delta$) distances are give in astronomical units (AU). The phase angles, $\alpha$, are given in degrees.  $1\sigma$ errors are given. 

\begin{figure}[htp]
\plotone{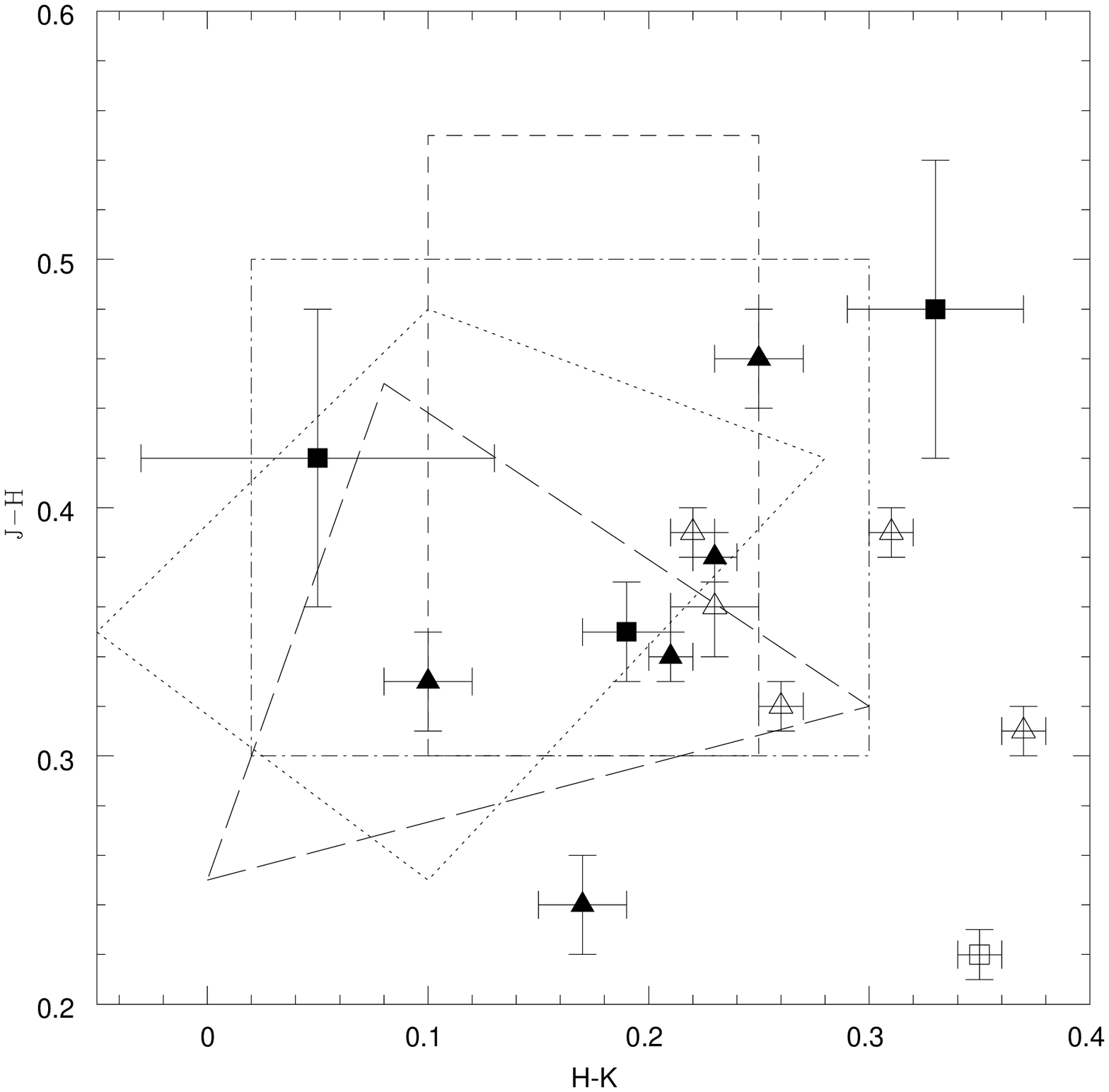}
   \caption{Grav \& Holman, Near-Infrared Photometry of Irregular Satellites of Jupiter and Saturn }
   \label{fig:jhk}
\end{figure}

{\bf Figure \ref{fig:jhk}.} Here the $H-K_s$ vs. $J-H$ near-infrared colors are shown for all the objects observed in this paper. The open and filled points indicates the optical color (grey and light red, respectively) of the irregular satellites as found by \cite{Grav.2003b}. Given are $1\sigma$ errorbars. The four areas sketched show the approximate phase space covered by different asteroidal classes as observed by the 2MASS projects. Plotted are the C- (dotted), F- (long dashed), D- (short dashed) and P-type (dotted-short dashed) asteroidal classes \citep{Sykes.2000b}.

\begin{figure*}[htp]
   \centering
\plotone{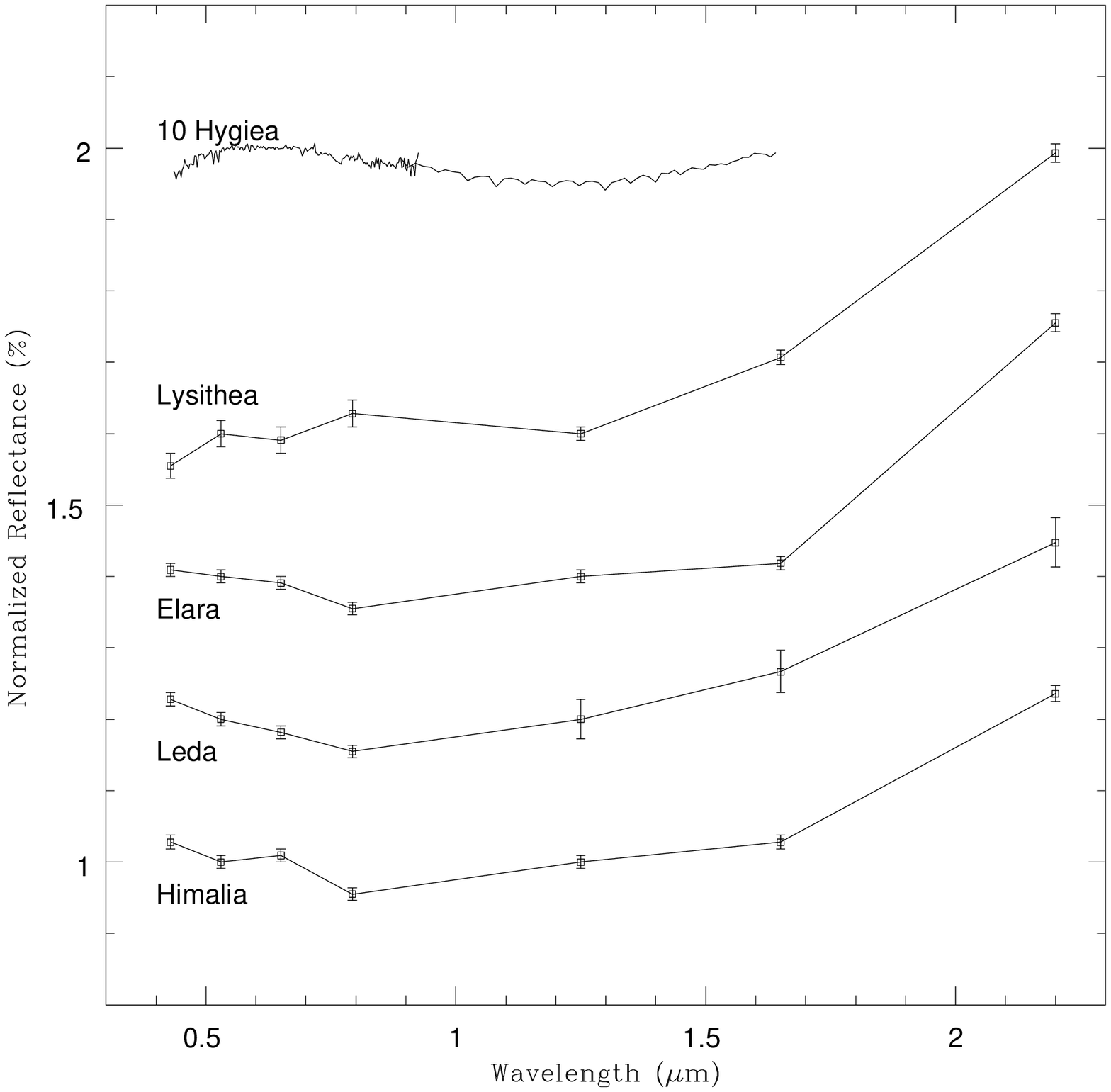}
   \caption{Grav \& Holman, Near-Infrared Photometry of Irregular Satellites of Jupiter and Saturn}
   \label{fig:him_spectra}
\end{figure*}

\begin{figure*}[htp]
   \centering
\plotone{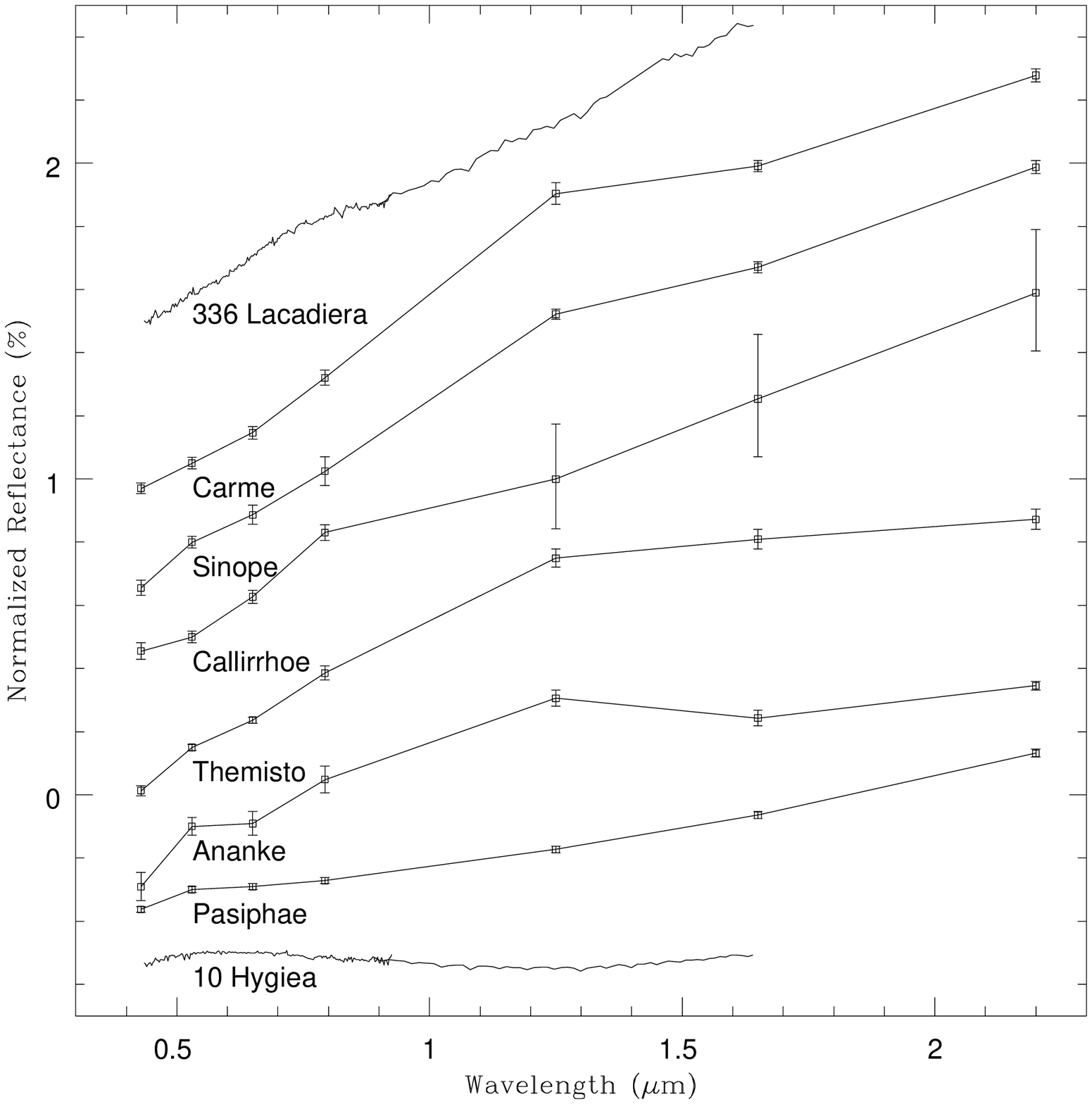}
   \caption{Grav \& Holman, Near-Infrared Photometry of Irregular Satellites of Jupiter and Saturn}
   \label{fig:jup_spectra}
\end{figure*}

{\bf Figure \ref{fig:him_spectra}.} Same as figure \ref{fig:jup_spectra}, but showing the four members of the Himalia-family. Note the subtle difference in the optical of J~XI~Lysithea from the rest. The asteroids 10 Hygeia (C-type)  is shown for comparison \citep{Bus.2002,Burbine.2002}.

{\bf Figure \ref{fig:jup_spectra}.} Shown are the normalized reflectance spectra of the Jovian irregular satellite (excluding the four members of the Himalia-family, which are shown in figure \ref{fig:him_spectra}). All spectra are normalized at the $V$ filter and all, but the spectrum of Sinope, have been offset for better comparison. The asteroids 10 Hygeia (C-type) and 336 Lacadiera (D-type) are shown for comparison \citep{Bus.2002,Burbine.2002}.

\begin{figure*}[htp]
   \centering
\plotone{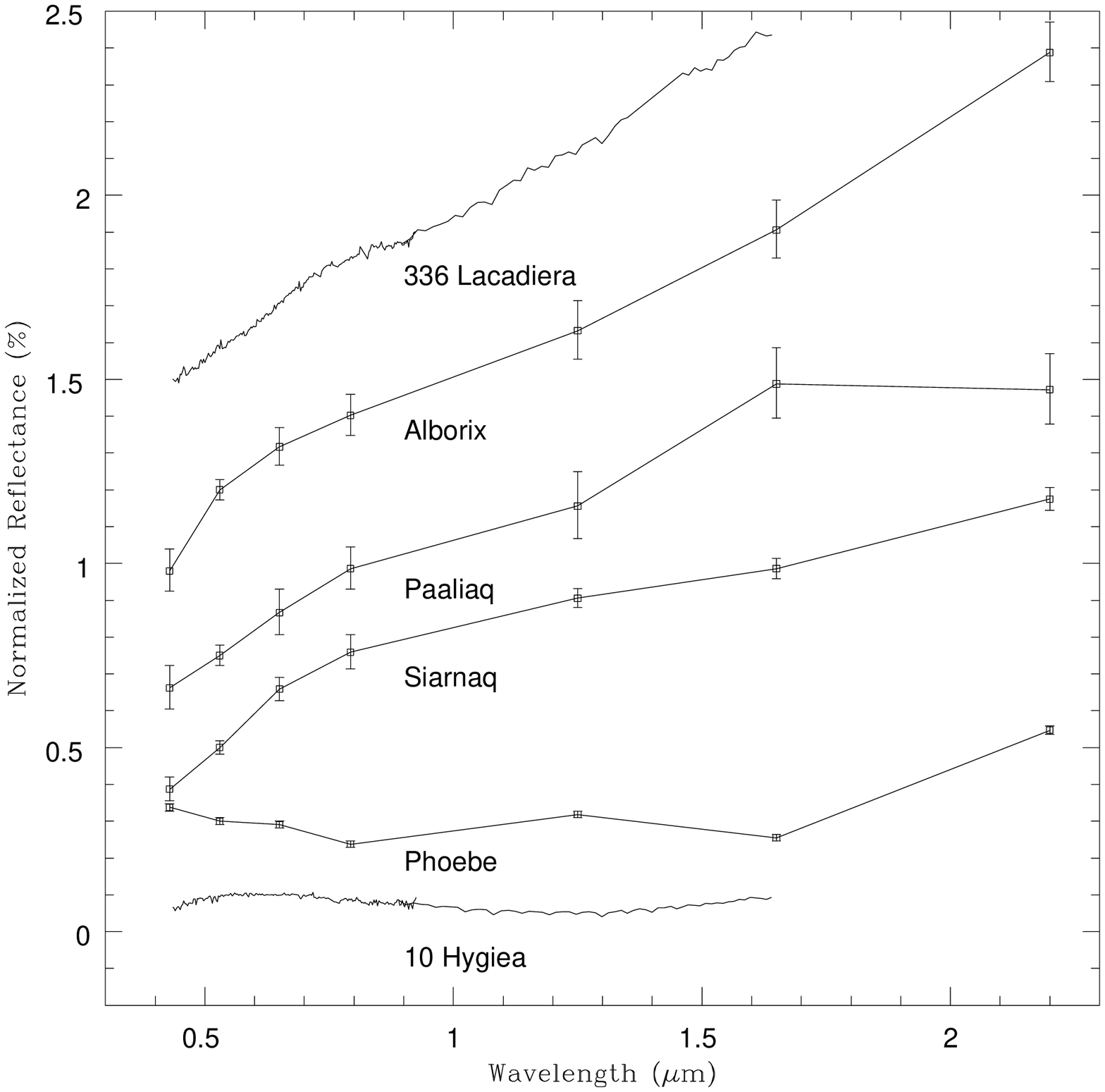}
   \caption{Grav \& Holman, Near-Infrared Photometry of Irregular Satellites of Jupiter and Saturn}
   \label{fig:sat_spectra}
\end{figure*}

{\bf Figure \ref{fig:sat_spectra}.} Same as figure \ref{fig:jup_spectra}, but showing the four Saturnian irregular satellites observed in this paper. Note the striking similarities of S~XX~Paaliaq and S~XXIX~Siarnaq. The asteroids 10 Hygeia (C-type) and 336 Lacadiera (D-type) are shown for comparison \citep{Bus.2002,Burbine.2002}.

%

\bibliography{ms}

\clearpage
\thispagestyle{empty}
\begin{sidewaystable*}[htp]
\begin{center}
\begin{tabular}{lccccccccccc}
 Object & Date (UT) && Airmass & $r$ & $\Delta$ & $\alpha$ 
              && $K_s$ & $J-H$ & $H-K_s$ & $J-K_s$ \\
\hline
S~IX~Phoebe & 2003 Feb. 18th 08:55 && 1.3-1.4 & 9.052 & 8.614 & 5.745 
                && $14.86\pm0.01$ & $0.22\pm0.01$ & $0.35\pm0.01$ & $0.57\pm0.01$\\
S~XXIX~Siarnaq & 2003 Feb. 18th 06:00 && 1.0-1.1 & 8.983 & 8.556 & 5.822 
                && $18.49\pm0.01$ & $0.35\pm0.02$ & $0.19\pm0.02$ & $0.54\pm0.02$ \\
S~XX~Paaliaq & 2003 Feb. 20th 06:20 && 1.0-1.1 & 9.114 & 8.721 & 5.827 
               && $19.43\pm0.06$ & $0.42\pm0.06$ & $0.05\pm0.08$ & $0.47\pm0.09$ \\
S~XXVI~Albiorix & 2003 Feb. 19th 05:15 && 1.0-1.1 & 9.076 & 8.667 & 5.814 
               && $18.79\pm0.03$ & $0.48\pm0.06$ & $0.33\pm0.04$ & $0.81\pm0.07$\\
\\	     
J~VI~Himalia    & 2003 Feb. 20th 09:50 && 1.0-1.1 & 5.254 & 4.315 & 3.727 
                   && $13.35\pm0.01$ & $0.32\pm0.01$ & $0.26\pm0.01$ & $0.58\pm0.01$ \\
J~VII~Elara         & 2003 Feb. 20th 09:25 && 1.0-1.1 & 5.308 & 4.376 & 3.911 
                   && $15.27\pm0.01$ & $0.31\pm0.01$ & $0.37\pm0.01$ & $0.68\pm0.01$ \\
J~VIII~Pasiphae & 2003 Feb. 19th 08:50 && 1.0-1.1 & 5.203 & 4.269 & 3.932 
                   && $15.10\pm0.01$ & $0.39\pm0.01$ & $0.22\pm0.01$ & $0.61\pm0.01$ \\
J~IX~Carme      & 2003 Feb. 20th 12:00 && 1.3-1.4 & 5.196 & 4.264 & 4.036 
                   && $15.74\pm0.01$ & $0.34\pm0.01$ & $0.21\pm0.01$ & $0.55\pm0.02$ \\
J~X~Sinope     & 2003 Feb. 19th 08:15 && 1.0-1.1 & 5.194 & 4.254 & 3.721 
                   && $16.05\pm0.01$ & $0.38\pm0.01$ & $0.23\pm0.01$ & $0.61\pm0.01$ \\
J~XI~Lysithea   & 2003 Feb. 20th 08:45 && 1.0-1.1 & 5.366 & 4.425 & 3.538 
                   && $16.30\pm0.01$ & $0.39\pm0.01$ & $0.31\pm0.01$ & $0.70\pm0.01$ \\
J~XII~Ananke    & 2003 Feb. 18th 13:00 && 1.6-2.0 & 5.443 & 4.502 & 3.493 
                   && $17.33\pm0.01$ & $0.24\pm0.02$ & $0.14\pm0.02$ & $0.38\pm0.02$ \\
                  & 2003 Feb. 20th 12:30 && 1.4-1.7 & 5.442 & 4.512 & 3.875 
                   && $17.32\pm0.02$ & $0.25\pm0.02$ & $0.20\pm0.03$ & $0.45\pm0.03$ \\
J~XIII~Leda         & 2003 Feb. 19th 09:20 && 1.0-1.1 & 5.253 & 4.308 & 3.493 
                   && $18.09\pm0.03$ & $0.36\pm0.02$ & $0.23\pm0.03$ & $0.59\pm0.03$ \\ 
J~XVIII~Themisto  & 2003 Feb. 18th 09:40 && 1.0-1.1 & 5.283 & 4.335 & 2.322 
                    && $18.30\pm0.02$ & $0.33\pm0.02$ & $0.10\pm0.02$ & $0.43\pm0.02$ \\
J~XVII~Callirrhoe & 2003 Feb. 19th 11:25 && 1.1-1.5 & 5.231 & 4.277 & 3.861 
                    && $19.17\pm0.10$ & $0.46\pm0.07$ & $0.25\pm0.12$ & $0.71\pm0.12$ \\
\end{tabular}
  \caption[ND] {Grav \& Holman, Near-Infrared Photometry of Irregular Satellites of Jupiter and Saturn}
  \label{tab:obs} 
\end{center}
\end{sidewaystable*}

\clearpage


\end{document}